\def\pr{0}
\documentclass[a4paper,12pt, epsfig]{article}
\usepackage{epsfig}
\usepackage{epstopdf}
\usepackage{graphicx}
\usepackage{ifthen}

\usepackage{amsmath}
\usepackage[psamsfonts]{amssymb}
\usepackage{euscript}

\usepackage{latexsym}

\newskip\humongous \humongous=0pt plus 1000pt minus 1000pt

\newif\ifdtup

\def\hsp{,\hspace{.7cm}}

\ifthenelse{\equal{\pr}{1}}{

\renewcommand{\cos}{\textrm{cos}}
\renewcommand{\sin}{\textrm{sin}}

\renewcommand{\(}{\begin{equation}}
\renewcommand{\)}{end{equation} \vspace{-.05in}\linebreak}

\newcounter{saveeqn}
\newcounter{savealpheqn}

\newcommand{\alpheqn}{\setcounter{saveeqn}{\value{equation}}%
  \stepcounter{saveeqn}\setcounter{equation}{0}%
  \renewcommand{\theequation}{\mbox{\arabic{section}.\arabic{saveeqn}
\alph{equation}}}
  \renewcommand{\)}{\end{equation}}}
\def\part#1{\frac{\partial}{\partial{#1}}}%
\def\group#1{\refstepcounter{equation}\setcounter{saveeqn}
 {\value{equation}}%
  \label{#1}\setcounter{equation}{0}%
\renewcommand{\theequation}{\mbox{\arabic{section}.\arabic{saveeqn}
\alph{equation}}}
  \renewcommand{\)}{\end{equation}}}
\newcommand{\reseteqn}{\setcounter{equation}{\value{saveeqn}}%
  \renewcommand{\theequation}{\arabic{section}.\arabic{equation}}%
  \renewcommand{\)}{\end{equation}}}

\newcommand{\aalpheqn}{\setcounter{saveeqn}{\value{equation}}%
  \stepcounter{saveeqn}\setcounter{equation}{0}%
  \renewcommand{\theequation}{\mbox{
        \Alph{subsection}.\arabic{saveeqn}\alph{equation}}}
   \renewcommand{\)}{\end{equation}}}
\newcommand{\areseteqn}{\setcounter{equation}{\value{saveeqn}}%
  \renewcommand{\theequation}{\Alph{subsection}.\arabic{equation}}%
  \renewcommand{\)}{\end{equation}}}

\renewcommand{\thefootnote}{\alph{footnote}}
\renewcommand{\(}{\begin{equation}}
\renewcommand{\)}{\end{equation}}
\newcommand{\ba}{\begin{eqnarray}}
\newcommand{\ea}{\end{eqnarray}}

\newcommand{\bp}{\mathop{\vtop{\ialign{##\crcr
   $\hfil\displaystyle{}\hfil$\crcr\noalign{\kern-13pt\nointerlineskip}
   \BIG{(}\hskip0pt\crcr\noalign{\kern3pt}}}}}
\newcommand{\cbp}{\mathop{\vtop{\ialign{##\crcr
   $\hfil\displaystyle{}\hfil$\crcr\noalign{\kern-13pt\nointerlineskip}
   \BIG{)}\hskip0pt\crcr\noalign{\kern3pt}}}}}
\newcommand{\pa}{\mathop{\vtop{\ialign{##\crcr
    
$\hfil\displaystyle{\oplus}\hfil$\crcr\noalign{\kern+1pt\nointerlineskip 
}
   \hspace{.08in}$^{\alpha=0}$\hskip6pt\crcr\noalign{\kern3pt}}}}}
\renewcommand{\hsp}{,\hspace{.3in}}

\numberwithin{equation}{section}
\renewcommand{\theequation}{\mbox{\arabic{equation}}}

\catcode`\@=11
\def\vereq#1#2{\lower3pt\vbox{\baselineskip1.5pt \lineskip1.5pt
\ialign{$\m@th#1\hfill##\hfil$\crcr#2\crcr\sim\crcr}}}
\catcode`\@=12

\renewcommand{\(}{\begin{equation}}
\renewcommand{\)}{\end{equation}}

\newcommand{\beas}{\begin{eqnarray*}}
\newcommand{\eeas}{\end{eqnarray*}}

\newcommand{\bquo}{\begin{quote}}
\newcommand{\enqu}{\end{quote}}

}{}

\newcommand{\C}{{\mathbb C}}
\newcommand{\cp}{{\mathrm{\mathbb CP}}}

\newcommand{\psip}{\psi^\prime}
\renewcommand{\sup}{\sqrt{1-\psi^{\prime 2}}}
\newcommand{\sz}{{\rm{sign}}(z)}
\newcommand{\zsz}{\left(z-z_0\sz\right)}
\newcommand{\dist}{x^2+y^2+\zsz^2}

\newcommand{\beq}{\begin{equation}}
\newcommand{\eeq}{\end{equation}}
\newcommand{\bea}{\begin{eqnarray}}
\newcommand{\eea}{\end{eqnarray}}

\newskip\humongous \humongous=0pt plus 1000pt minus 1000pt

\newif\ifdtup

\def\noprl#1{\ifthenelse{\equal{\pr}{1}}{}{#1} }

\catcode`\@=11

\@addtoreset{equation}{section}
\def\theequation{\thesection.\arabic{equation}}

\def\@normalsize{\@setsize\normalsize{15pt}\xiipt\@xiipt
\abovedisplayskip 14pt plus3pt minus3pt%
\belowdisplayskip \abovedisplayskip
\abovedisplayshortskip \z@ plus3pt%
\belowdisplayshortskip 7pt plus3.5pt minus0pt}

\def\small{\@setsize\small{13.6pt}\xipt\@xipt
\abovedisplayskip 13pt plus3pt minus3pt%
\belowdisplayskip \abovedisplayskip
\abovedisplayshortskip \z@ plus3pt%
\belowdisplayshortskip 7pt plus3.5pt minus0pt
\def\@listi{\parsep 4.5pt plus 2pt minus 1pt
      \itemsep \parsep
      \topsep 9pt plus 3pt minus 3pt}}

\relax

\catcode`@=12

\catcode`\@=11

\def\section{\@startsection{section}{1}{\z@}{3.5ex plus 1ex minus
    .2ex}{2.3ex plus .2ex}{\large\bf}}

\def\thesection{\arabic{section}}
\def\thesubsection{\arabic{section}.\arabic{subsection}}

\def\appendix{\setcounter{section}{0}
  \def\thesection{Appendix \Alph{section}}
  \def\thesubsection{\Alph{section}.\arabic{subsection}}
  \def\theequation{\Alph{section}.\arabic{equation}}}

\begin{document}
\def\thefootnote{\fnsymbol{footnote}}
\def\thetitle{Spiked Monopoles}
\def\autone{Jarah Evslin}
\def\auttwo{Baiyang Zhang}
\def\affa{Institute of Modern Physics, NanChangLu 509, Lanzhou 730000, China}
\def\affb{University of the Chinese Academy of Sciences, YuQuanLu 19A, Beijing 100049, China}

\ifthenelse{\equal{\pr}{1}}{
\title{\thetitle}
\author{\autone}
\affiliation {\affa}
\affiliation {\affb}
}{
\begin{center}
{\large {\bf \thetitle}}

\bigskip

\bigskip

{\large \noindent  \autone \footnote{jarah@impcas.ac.cn}}

\vskip.7cm

1) \affa\\
2) \affb\\

\end{center}
}

\begin{abstract}
\noindent

\noindent
We introduce the spiked monopole, which is a 't Hooft-Polyakov monopole with two charged scalar Higgs fields, of which one enjoys a quartic self-interaction.  The free Higgs field behaves as in a BPS monopole, reducing the inter-monopole repulsion.  The other Higgs has a spiked profile similar to a non-BPS monopole.  Using the methods from numerical relativity recently adapted to the Yang-Mills-Higgs theory by Vachaspati, we simulate the interactions of such monopoles.  During the long lifetime of these simulations the individual monopoles are stable.  We find that they are always repulsive, with a small repulsion only when the interaction Higgs VEV is proportionately small.  We briefly comment on implications for giant monopole dark matter models and on supermassive black hole seeding by the spikes.

\end{abstract}

%
\setcounter{footnote}{0}
\renewcommand{\thefootnote}{\arabic{footnote}}


\ifthenelse{\equal{\pr}{1}}{
\maketitle
}{}


\section{Introduction}

Halo-sized non-BPS 't Hooft-Polyakov monopole dark matter models \cite{bjarkedark,procdark} predict dark matter halos with density distributions which are the energy distributions of the corresponding classical field theory solutions.  In other words, they are automatically cored and pseudo-isothermal in the sense that at intermediate radii their density falls as the inverse squared radius, resolving the core/cusp problem \cite{moorecc}.  Moreover, they are described by a single parameter corresponding to their magnetic charge, reproducing the observed one-parameter universality of rotation curves in spiral galaxies \cite{salucci}.  Dirac quantization also ensures a minimum mass, potentially resolving the missing satellites problem \cite{klypin,moore}.  The main phenomenological obstruction to such dark matter models is that the monopoles repel, unlike real dark matter halos whose long distance interactions are gravitationally dominated.  Various proposed solutions to this problem have been proposed, from screening by light antimonopoles of another flavor in the original references to confinement inside of Skyrmions in Ref.~\cite{q2}.  The first mechanism has yet to be realized in a concrete model\footnote{However in Ref.~\cite{superalex}, in a related context, the dark abelian gauge field is screened by charged dark matter.  The electromagnetic dual of that model exhibits the desired magnetic screening.} while the second leads to metastability, not stability, for large halos.  

In the current note we investigate another potential solution.  Our monopoles need to be deeply non-BPS to exhibit the desired isothermal density profile in the inner region, yet at long distances we would like the cancellation of forces characteristic of BPS monopoles.  We will attempt to achieve the best of both worlds by including two charged scalar Higgs fields in our theory, one of which has a large self-interaction and so is deeply non-BPS, yielding the pseudoisothermal spike in the core, while the second has no self-interaction and so serves to cancel the magnetic repulsion at large distances.

Even if the long distance interaction can be made negligibly small for static monopoles at a fixed separation, there is no guarantee that the interactions between monopoles will be sufficiently small to satisfy all observational bounds.  BPS monopoles have very large short distance interactions and also long distance interactions proportional to the relative velocity squared \cite{gm}.  In addition, the phenomenological bounds themselves, derived from cluster interactions, are still quite controversial \cite{wittmanbound1,wittmanbound2}.  These bounds are derived using very simple models of repulsion by a central potential yielding scattering at a fixed angle, which is quite different from the velocity dependent, non-central interactions characteristic of BPS monopoles.  Indeed, the interactions of multimonopole systems with one another, even in the BPS case, depends on an understanding of the internal kinematics of each system, which again is dominated by such non-central interactions, and so is nontrivial.  We will return to this problem in a sequel.

Generally speaking, solitonic dark matter models fall into two categories.  First, each dark matter halo may consist of a single soliton, albeit of high charge.  Such halos presumably formed from a merger of charge one solitons which needed to be light enough to avoid introducing too much shot noise in the matter power spectrum \cite{afshordishot}.  Such halos, if cool enough, will have a shape determined by the profile of the soliton solution and so will yield universal halo profiles.  The other possibility is that each halo consists of a number of solitons which move sufficiently quickly to form a dispersion supported structure.  In this case the individual soliton subhalos must satisfy the bound \cite{afshordishot}.

In our study below, we will find that acceptably small repulsion requires the spike to have much less mass than the total halo\footnote{As a result kinematic bounds such as that in Ref.~\cite{spikebound} will be easily satisfied.}.  Thus although the isothermal profile of the spike is tantalizingly similar to observed halo plus baryon density profiles, yielding flat rotation curves for example, it seems unlikely that the spike can contain a large enough fraction of the halo mass to agree with observations.  Therefore, if realized in Nature the spiked monopole scenario would likely fall under the second category above.  The universality of halo shapes would therefore not be a direct consequence of the soliton solution.  In this case, the monopole gas may be thin enough to satisfy upper limits on the dark matter scattering cross section, although this may require individual monopoles which are so small that the minimum mass alone would not yield a solution to the missing satellite problem.

On the bright side, the spikes in the soliton solution would necessarily form seeds for the formation of today's supermassive black holes.  With the discovery of supermassive black halos at ever larger redshift \cite{bhnature}, it has become ever more difficult to produce convincing scenarios of their growth \cite{volontieri}.  Large black holes require large seeds \cite{loebseed} or else seeds which were created very early \cite{primbh1,primbh2}.  The spikes of spiked monopoles, produced via the Kibble mechanism, would provide very early seeds.

\section{Individual Spiked Monopoles} \label{unosez}
\subsection{Einstein-Yang-Mills-Higgs}
Although in the present note we will only be interested in solutions in the range of parameter space in which gravity is essentially Newtonian, we are motivated in part by the formation of black holes and so it will eventually be useful to embed our solutions in general relativity.  Therefore we will introduce our monopoles in the context of Einstein gravity coupled to the SU(2) Yang-Mills Higgs theory, and later specialize to the case of Newtonian gravity.

Einstein-Yang-Mills-Higgs theory is defined by the following action
\bea
S&=&\int\sqrt{-{\rm{det}}(g)}\mathcal{L}\hsp
\mathcal{L}=\mathcal{L}_{\rm{grav}}+\mathcal{L}_{\rm{YM}}+\mathcal{L}_{\rm{H}}\\
\mathcal{L}_{\rm{grav}}&=&\frac{1}{4k} R\hsp k=4\pi G_N\\
\mathcal{L}_{\rm{YM}}&=&-\frac{1}{2g^2}{\rm{Tr}}\left(F_{\mu\nu}F^{\mu\nu}\right)\hsp
F_{\mu\nu}=\partial_\mu A_\nu-\partial_\nu A_\mu-i[A_\mu,A_\nu]\\
\mathcal{L}_{\rm{H}}&=&\sum_{I}\left[{\rm{Tr}}\left(D_\mu\Phi_I D^\mu\Phi_I\right)-\frac{\lambda_I}{4}\left(2{\rm{Tr}}(\Phi_I^2)-v_I^2\right)^2\right],\ \ 
D_\mu\Phi_I=\partial_\mu\Phi_I-i[A_\mu,\Phi_I].
\eea
The analogue of the BPS condition for $\phi_1$ is $\lambda_1=0$, which will be imposed in subsequent sections.  We will adopt the spherically symmetric, static Ansatz \cite{maison}
\bea
ds^2&=&\sigma^2(r) N(r) dt^2-\frac{dr^2}{N(r)^2}-r^2(d\theta^2+{\rm{sin}}^2(\theta)d\phi^2)\hsp
N(r)=1-\frac{2km(r)}{r}\\
A_i&=&\epsilon_{aik}\frac{x^k}{r^2}(1-w(r))T^a\hsp
\Phi_I=v_I \phi_I(r)\frac{x^j T^j}{r} \label{ans}
\eea
where for brevity we have mixed Cartesian $x^i$ and spherical $(r,\theta,\phi)$ coordinates.  We have also suppressed dependence on space-time while making dependence on $r$ alone explicit, to highlight that with this Ansatz the equations of motion become ordinary differential equations in $r$.  All pairs of indices are summed implicitly regardless of whether they are up or down except for the flavor index, which will always be denoted using capital letters.  The gauge generators are normalized such that ${\mathrm{Tr}}(T^aT^b)=\delta^{ab}/2$.

With this Ansatz, Einstein's Equations reduce to one constraint and one dynamical equation
\beq
G_{t}^t=2k T_t^t\hsp
G_t^t-G_r^r=2k(T_t^t-T_r^r).
\eeq
Multiplying the former by $r^2/2k$ and the latter by $r/2kN(r)$ one obtains the two equations
\bea
m^\prime(r)&=&\frac{N(r)w^{\prime 2}(r)}{g^2}+\frac{\left(1-w^2(r)\right)^2}{2g^2r^2}+\sum_{I}\left[v_I^2w^2(r)\phi_I^2(r)+\frac{\lambda_I v_I^4 r^2}{4}\left(\phi_I^2(r)-1\right)^2\right]\label{grav1}\\
\frac{\sigma^\prime(r)}{k\sigma(r)}&=&\frac{2w^{\prime 2}(r)}{g^2 r}+\sum_{I} v^2_I \phi_I^{\prime 2}(r) r.\label{grav2}
\eea
Further equations follow from the vanishing of the variation with respect to $A$ and $\Phi_I$ respectively
\bea
\left(N(r)\sigma(r)w^\prime(r)\right)^\prime&=&\frac{\sigma(r)}{r^2}w(r)\left(w^2(r)-1\right)+\sum_I g^2 v^2_I \phi^2_I(r) \sigma(r) w(r) \label{campo1}\\
\left(N(r)r^2 \sigma(r)\phi_I^\prime(r)\right)^\prime&=&2w^2(r)\sigma(r)\phi_I(r)+\lambda_I v^2_I r^2 \sigma(r)\left(\phi_I^2(r)-1\right)\phi_I(r). \label{campo2}
\eea

If $km(r)<<r$, which is generally the case for $v_I$ well below the Planck scale, then one may approximate $N(r)=\sigma(r)=1$.  In this case the two gravitational equations (\ref{grav1}) and (\ref{grav2}) can be ignored, as they can always be integrated to produce $N(r)$ and $\sigma(r)$ which are anyway approximated to be unity.  The remaining equations simplify to
\bea
w^{\prime\prime}(r)&=&\left[\frac{w^2(r)-1}{r^2}+\sum_I g^2 v^2_I \phi_I^2(r)\right]w(r) \label{weq}\\
\left(r^2\phi_I^\prime(r)\right)^\prime&=&\left[2w^2(r)+\lambda_I v^2_I r^2 (\phi^2_I(r)-1)\right]\phi_I(r).\label{feq}
\eea
The spiked monopole is the solution to the ordinary differential equations (\ref{weq}) and (\ref{feq}) with $\lambda_1=0$ and the boundary conditions
\beq
w(0)=1\hsp
w(\infty)=0\hsp
\phi_I(0)=0\hsp
\phi_I(\infty)=1. \label{bcond}
\eeq

\subsection{Asymptotics}
At large $r$, as the W boson is Higgsed, $w(r)$ exponentially goes to zero.  As it is massive when varied about the minimum of its potential, $\phi_2(r)$ also exponentially goes to $1$.  Thus at high $r$ the only non-exponentially suppressed evolution is that of $\phi_1(r)$.  Dropping the exponentially suppressed $w(r)$ term, it is described by (\ref{feq})
\beq
\left(r^2\phi_1^\prime(r)\right)^\prime=0
\eeq
whose solutions with the boundary conditions (\ref{bcond}) are
\beq
\phi_1=1-\frac{c}{r}. \label{asi}
\eeq
In the BPS case, corresponding here to $\lambda_2v_2=0$, one finds $c=1/gv$ and the attractive force caused by the scalar cancels the monopole's repulsive magnetic force.  However more generally $c$ appears to be unconstrained.

\subsection{Numerical Solutions}

We have numerically solved this system of ordinary differential equations for various values of $v_1$, $v_2$ and $\lambda_2$.  We have found that
\beq
0\leq c\leq 1/gv \label{cmax}
\eeq
where the upper bound is saturated only in the BPS case $\lambda_2v_2=0$.  In particular, the failure of $c$ to saturate its upper bound appears to be monotonic in $\lambda_2$ and $v_2$.

\begin{figure} 
\begin{center}
\includegraphics[width=4.2in,height=2.5in]{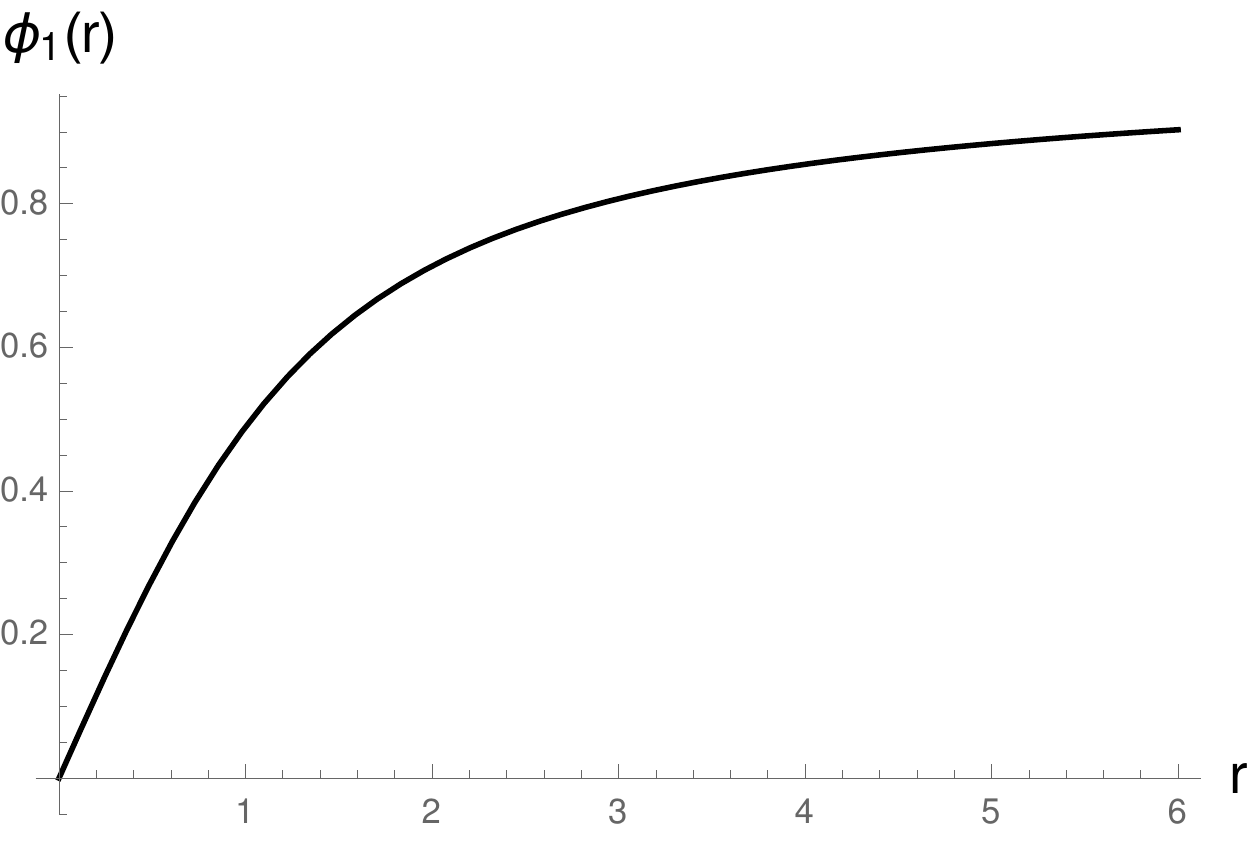}
\includegraphics[width=4.2in,height=2.5in]{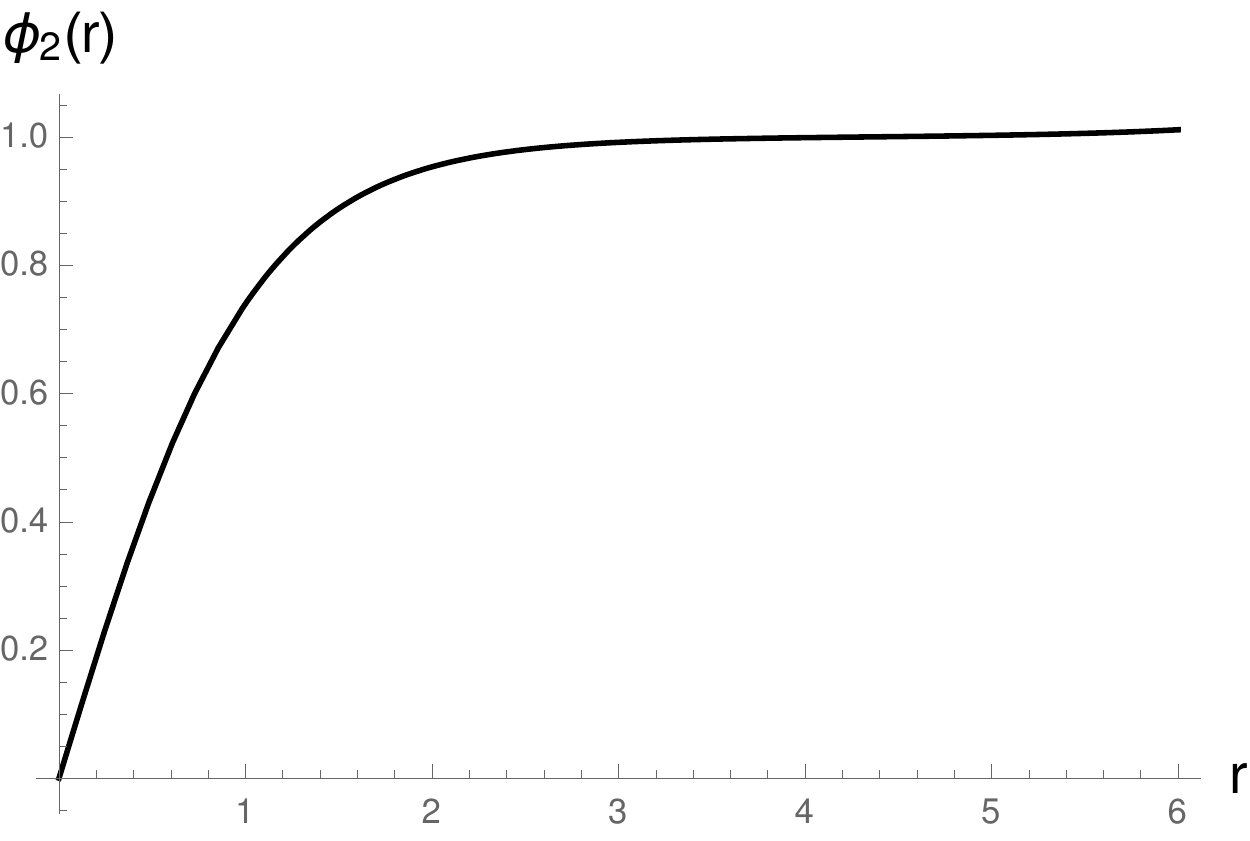}
\includegraphics[width=4.2in,height=2.5in]{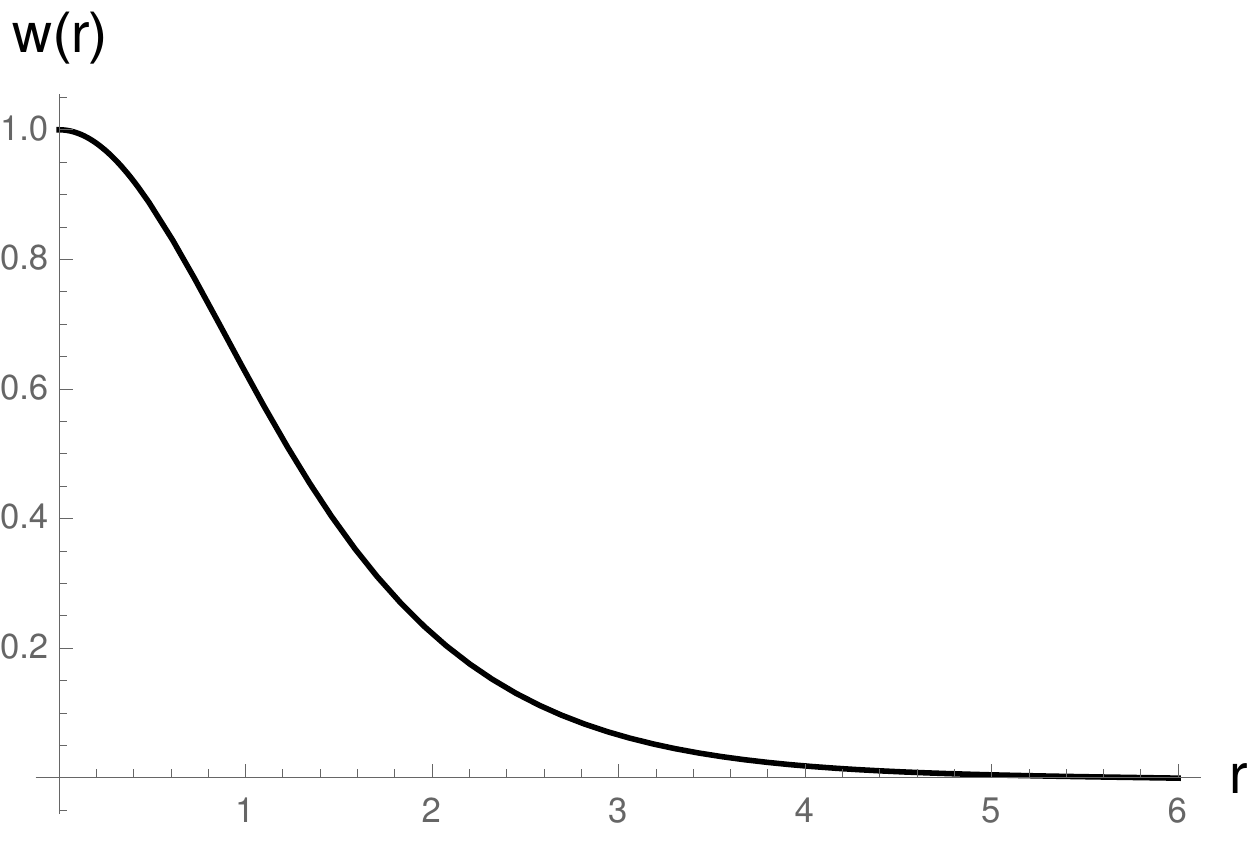}
\caption{In order from top to bottom, the functions $\phi_1(r)$, $\phi_2(r)$ and $w(r)$ are shown for $g=v_1=v_2=\lambda_2=1$. }
\label{1monfig}
\end{center}
\end{figure}

\begin{figure} 
\begin{center}
\includegraphics[width=4.2in,height=2.5in]{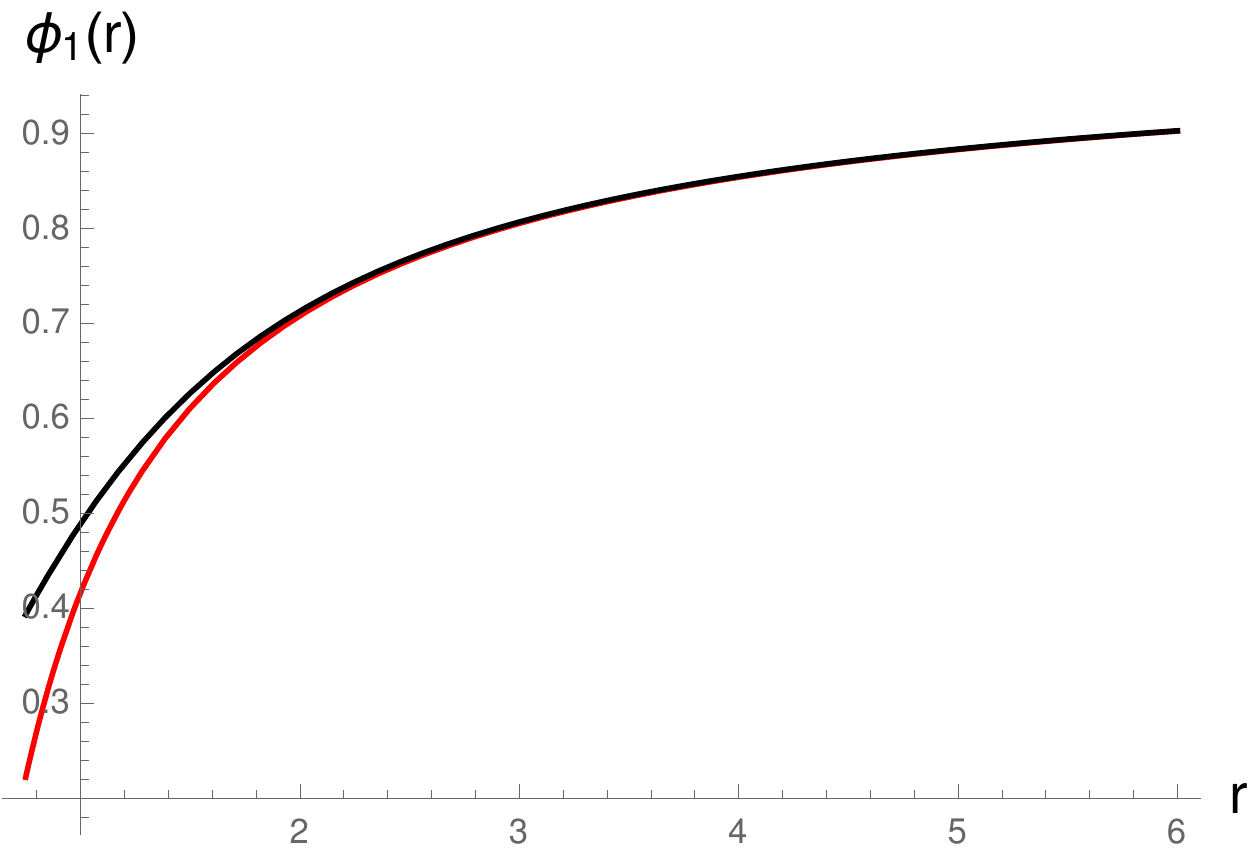}
\caption{The function $\phi_1(r)$ (black) is compared with $1-0.5835/r$ (red).}
\label{phi1fig}
\end{center}
\end{figure}

In the case $g=v_1=v_2=\lambda_2=1$, the functions $\phi_i(r)$ and $w(r)$ are drawn in Fig.~\ref{1monfig}.  In Fig.~\ref{phi1fig} we compare $\phi_1(r)$ with the asymptotic form (\ref{asi}) with $c=0.5835$, which is only about half of the BPS bound.  The agreement between these curves at high $r$ may lead one to suspect that, since $c$ does not saturate its BPS value, the scalar field $\phi_1$ in such monopoles is insufficient to balance the repulsive magnetic field, and so such monopoles repel.  We will see numerically that this is indeed the case.  We have also tried various nonrenormalizzable potentials for $\phi_2$ but were unable to violate the upper bound (\ref{cmax}), and so we expect that spiked monopoles will repel even in such cases.  

\section{Monopole Interactions}

We have found hedgehog-like spiked monopole solutions of the Yang-Mills-Higgs system.  In the rest of this paper, we will attempt to answer two questions regarding these solutions.

First of all, are they stable?  Even if they are stable against spherically-symmetric perturbations, this does not guarantee stability against all perturbations.  Topology guarantees that the winding number of each scalar field agrees with the magnetic charge, which is equal to unity.  However, both scalar fields interact with the gauge field and so it is not clear that such a simple concentric wrapping is the lowest energy solution.  One may worry, for example, whether the presence of multiple charged scalar fields could lead to the existence of semilocal solutions as happens in the case of strings \cite{semilocal,semilocalnoi}.

Second, do they really repel?  Sure, $c$ does not saturate its BPS bound.  But this is the bound necessary to cancel the magnetic repulsion in the case of a single scalar field.  Now there are two scalar fields.  This means that the nonabelian part of the magnetic field is confined to lower radii, which affects the distribution of the magnetic field although Gauss' law at large distances means that it cannot change the integrated flux.  In addition, if we are interested in the interactions of two spiked monopoles with one another, then the fact that both of them have modified scalar fields may lead to the possibility that the critical value of $c$ for force cancellation is not the BPS value.  After all, a larger fraction of the spiked monopole mass is in the form of scalar fields than in the case of a BPS monopole, and so perhaps the scalar field's attraction is somehow more powerful in this case, canceling the effect of the submaximal $c$?  To respond robustly to this question, we will simulate the evolution of such systems.

\subsection{Vachaspati's Numerical Method}

The Ansatz used above to obtain the spiked monopole solution used the temporal gauge $A_0=0$.  To numerically evolve the Yang-Mills-Higgs system, we need to write the evolution equations in a form that preserves this gauge choice.  

Our static, spiked monopole solution, as a solution of the full equations of motion, solves the Gauss constraints which arise from the variation with respect to $A_0$
\beq
\partial_k F_{0k}^a+\epsilon_{abc} A_k^b F_{0k}^c+\sum_I g^2 \epsilon_{abc} \Phi_I^b \dot\Phi_I^c=0 \label{vinc}
\eeq
even though $A_0$ is set to zero in the Ansatz.   Gauge-invariance guarantees that once the constraints are satisfied in the initial conditions, evolution under the (hyperbolic) equations with two time derivatives will continue to satisfy these constraints.   However on the lattice, imprecision due to the finite lattice spacing will cause violations of the constraints which, once present, are grown exponentially by the second order evolution equations.  Thus one needs a method of enforcing these constraints as the system evolves.  Solving the elliptic equations (\ref{vinc}) at each iteration would be quite time consuming.

Fortunately, Vachaspati has imported a method from numerical relativity for this purpose~\cite{vachrel}.  This method decomposes the second order Yang-Mills-Higgs evolution equations into first order equations and introduces an auxiliary, su(2) Lie algebra valued field $\Gamma^a$ whose evolution enforces these constraints.  In our case, the evolution equations are
\bea
\partial_t\phi_I^a&=&\dot\phi_I^a\\
\partial_t A_k^a&=&F_{0k}^a\\
\partial_t\dot\phi_I^a&=&\partial_k^2 \phi^a_I-2\epsilon_{abc}\partial_k\phi^b_IA_k^c+A_k^a\phi_I^bA_k^b-\phi_I^aA_k^bA_k^b-\lambda_I(\phi^b_I\phi^b_I-v_I^2)\phi_I^a+\epsilon_{abc}\Gamma^b\phi_I^c\\
\partial_t F_{0k}^a&=&\partial_j^2A_k^a+2\epsilon_{abc}A_j^b\partial_jA_k^c-\epsilon_{abc}A_j^b\partial_kA_j^c+A_j^aA_j^bA_k^b-A_k^aA_j^bA_j^b-\partial_k\Gamma^a\\
&&-\epsilon_{abc}A_k^b\Gamma^c+g^2\sum_I\left[-\epsilon_{abc}\phi_I^b\partial_k\phi_I^c- \phi_I^b\phi_I^bA_k^a+\phi^b_IA_k^b\phi_I^a\right]\\
\partial_t\Gamma^a&=&\partial_k F_{0k}^a-g_p^2\left[\partial_k F_{0k}^a+\epsilon_{abc} A_k^b F_{0k}^c+\sum_I g^2 \epsilon_{abc} \Phi_I^b \dot\Phi_I^c \right] \label{gev}
\eea
where $g_p$ is any constant.  In the continuum limit, the choice of $g_p$ is irrelevant because it multiplies the vanishing constraint (\ref{vinc}).  However on the lattice this constraint will fail to be zero as a result of numerical imprecision and so $g_p$ can be chosen at will to enforce stability.  Following Ref.~\cite{vachrel} we choose $g_p^2=0.75$, although unlike that reference we also set $g=1$.

The auxilliary field $\Gamma^a$ is given the initial value
\beq
\Gamma^a(t=0)=\partial_kA^a_k(t=0).
\eeq
Note that if the constraint is satisfied, the evolution equation (\ref{gev}) guarantees that $\Gamma$ will be equal to $\partial_k A_k$ at all times.  In practice, the failure of the constraint will lead to a deviation of $\Gamma$ which will push the solution back towards the constraint surface.

Simply discretizing time and evolving according to finite differences given by the evolution equations, numerical imprecisions grow exponentially and the configuration soon diverges.  To eliminate this problem, again following Ref.~\cite{vachrel}, we evolve using the second order Crank-Nicholson method.  This was shown to be the optimal order in Ref.~\cite{cnopt}.

\subsection{Covariant Absorbing Boundary Conditions}

Ref.~\cite{vachrel} introduced a new kind of absorbing boundary condition, in which the Laplacian is replaced with $\partial_r\partial_t$ to effectively make free waves travel outwards at the speed of light.  This boundary condition is not gauge covariant, as ordinary instead of covariant derivatives are employed.  This caused little problem for the authors as they considered only massive fields whose values were anyway exponentially suppressed near the boundary.  In our case, the field $\phi_1$ only decreases as $1/r$ and so its derivative at the boundary is not negligible.  As a result, when we attempted to use this kind of boundary condition, the constraints were violently violated near the boundary and this violation soon spread to the entire lattice.

Therefore we have instead introduced covariant boundary conditions.  We derived these boundary conditions by altering the metric at the boundary to
\beq
g_{tt}=g_{tr}=a\hsp g_{tk}=g_{rr}=g_{rk}=0\hsp g_{jk}=-b\delta_{jk}\hsp b<<a
\eeq
where $j$ and $k$ are perpendicular to $r$.  As the boundary conditions came from a modification of the metric, gauge-covariance is guaranteed.  Returning to a Minkowski metric without changing the form of the evolution equations, these boundary conditions change the evolution equations at the boundary to
\bea
\partial_t\dot\phi_I&=&-\partial_r\dot\phi_I-\epsilon_{abc} A_r^b \dot\phi_I^c-\frac{1}{2}F_{0r}^b\phi^c_I\\  
\partial_tF_{0k}^a&=&-\partial_r F_{0k}^a+\partial_k F_{0r}^a-\epsilon_{abc} A_r^bF_{0k}^c-\frac{1}{2}\epsilon_{abc}F_{0r}^bA_k^c.
\eea
Here one can recognize the first terms on the right hand sides as the boundary conditions of Ref.~\cite{vachrel} while the later terms serve to render them gauge covariant.

\subsection{Two-Monopole Initial Conditions}

We are interested in interactions of 2 spiked monopoles with each other.  In Sec.~\ref{unosez} we described the construction of a single spiked monopole.  This is a time-independent solution to the equations of motion.  It is not known if there are any time-independent solutions with two monopoles and in fact the repulsion that we will find makes the existence of such a solution unlikely, at least in the absence of gravity.  Therefore, since time-independence is out of the question, the choice of initial conditions for a simulation of 2 monopoles is somewhat arbitrary.

We will be guided by the following argument.  The Ansatz for $A_i$ in Eq.~(\ref{ans}) decomposes the gauge field into two terms.  The $1$ in the $(1-w(r))$ represents the long range abelian part of the field, whereas the $w(r)$ represents the W boson, which is massive outside of the core and exponentially falls to zero.  In the case at hand, we are interested in well separated monopoles.  So nearby one, it should resemble a single spiked monopole.  

With this vague motivation, we will place the monopoles at $x=y=0$, $z=\pm z_0$ with $z_0>0$ and we will divide the space into two regions along the plane $z=0$.  Each region contains one monopole and in each the $w$ term in (\ref{ans}) will be simply that corresponding to a single spiked monopole in that space.  It will not be differentiable at $z=0$, but it is already exponentially suppressed there.   The other term, essential for the asymptotic behavior of the 2-monopole system, will be taken from Manton's construction in Ref.~\cite{manton77}.

Repeating Manton's construction, to determine the fields at a point $p$ first one determines the angle $\theta_i$ between the $z$ axis and the line from $p$ to each monopole.  Their sum is the stream function
\beq
\psi={\rm{sin}}(\theta_1)+{\rm{sin}}(\theta_2)\hsp
\psi^\prime=\psi+{\rm{sign}}(z)\hsp {\rm{sign}}(z)=\frac{z}{|z|}.
\eeq
Let $\hat{\phi}$ be the unit vector in color space corresponding to $\phi_I$ at the point $p$.  We are adapting Manton's analysis in which there is only one scalar field, so this direction will necessarily be independent of the flavor index $I$, although the radial dependence does depend on the flavor
\beq
\Phi_I=v_I \phi_I(\tilde{r}) \hat{\phi}(\tilde{r}). \label{phid}
\eeq
Here $\tilde{r}$ is the distance to the nearest monopole, so the derivative will be discontinuous at $z=0$.  In our initial condition, we fix $\hat{\phi}$ to the large $\tilde{r}$ form of Ref.~\cite{manton77}
\beq
\hat{\phi}=\sqrt{1-\psi^{\prime 2}}\frac{x}{\sqrt{x^2+y^2}}T^1+{\rm{sign}}(z)\sqrt{1-\psi^{\prime 2}}\frac{y}{\sqrt{x^2+y^2}}T^2+{\rm{sign}}(z)\psi^\prime T^3
\eeq
while $\phi_I(\tilde{r})$ are taken to be the single monopole solutions.

At large $\tilde{r}$ the gauge field only depends on $\hat{\phi}$
\beq
A^{\rm{asy}}_k=[\partial_k\hat\phi,\hat\phi].
\eeq
In the case of a single monopole, $A^{\rm{asy}}$ is in fact just the abelian term in Eq.~(\ref{ans}).  Therefore the gauge field of a single monopole is just the sum of $A^{\rm{asy}}$ and the $w$ term in (\ref{ans}).  As we would like the two monopole initial condition to reproduce the single monopole solution near each monopole, our initial condition will be that on each side of $z=0$ the gauge field is just this sum
\beq
A_i=A_i^{\rm{asy}}-\epsilon_{ika}\frac{w(\tilde{r}) \tilde{x}^k}{\tilde{r}^2}T^a.
\eeq
Here $\tilde{r}$ is the distance to the nearest monopole, and $\tilde{x}^k$ is the coordinate centered on the nearest monopole.  Recalling that $\Phi_I$ is given by (\ref{phid}) with $\phi_I$ given by the one monopole solution for the nearest monopole, the initial conditions for the 2 monopole configurations are now determined.

Converting to global Cartesian coordinates with $\rho=\sqrt{x^2+y^2}$, the distance to the nearest monopole is $\sqrt{\dist}$ and we can evaluate the initial conditions for $A$ in terms of $\psip$ and $w$
\bea
A_i&=&A_i^{\rm{asy}}+A_i^w\\
A_x^{\rm{asy}}&=&\left[-\frac{y}{\rho}\frac{\partial_x\psi^\prime}{\sup}-\frac{xy}{\rho^3}\psip\sup\right]T^1\\
&&+\sz\left[\frac{x}{\rho}\frac{\partial_x\psip}{\sup}-\frac{y^2}{\rho^3}\psip\sup\right]T^2+\sz \frac{y}{\rho^2}\left(1-\psi^{\prime 2}\right) T^3\nonumber\\
A_y^{\rm{asy}}&=&\left[-\frac{y}{\rho}\frac{\partial_y\psi^\prime}{\sup}+\frac{x^2}{\rho^3}\psip\sup\right]T^1\\
&&+\sz\left[\frac{x}{\rho}\frac{\partial_y\psip}{\sup}+\frac{xy}{\rho^3}\psip\sup\right]T^2-\sz \frac{x}{\rho^2}\left(1-\psi^{\prime 2}\right) T^3\nonumber\\
A_z^{\rm{asy}}&=&-\frac{\partial_z\psip}{\rho\sup}\left[yT^1-x\sz T^2\right]\\
A_x^w&=&\frac{w\zsz\sz}{\dist}T^2-\frac{w y\sz}{\dist}T^3\\
A_y^w&=&-\frac{w\zsz\sz}{\dist}T^1+\frac{w x\sz}{\dist}T^3\\
A_z^w&=&\frac{wy}{\dist}T^1-\frac{w x\sz}{\dist}T^2.
\eea
Here $w$ is always evaluated at $\sqrt{\dist}$, but for brevity the argument is left implicit.

\section{Results}

\subsection{Simulation Parameters}

We considered $g=1$.  We varied $v_1$, $v_2$ and $\lambda_2$ but performed most of our simulations on the combination $v_1=v_2=\lambda_2=1$.  We simulated single spiked monopoles and also pairs of spiked monopoles starting at initial positions $x=y=0$ and $z=\pm z_0$,  In the case of single monopoles, we ran simulations with various initial velocities.  To create a moving single monopole, we Lorentz transformed the initial conditions and then gauge transformed back to the temporal gauge $A_0=0$, as the evolution equations (\ref{gev}) are given in that gauge.  

We identified the monopole position with the zero of the field $\phi_I$ interpolated between the grid points.  We found that this definition of the position was in general quite independent of which flavor of the field was used.  In other words, even during the interactions we found that the zeros of the fields were separated by an amount consistent with the uncertainties of the simulations.

We used two overlapping grids of the same dimensions.  Each grid had dimensions of between 16 and 34 in $x$ and $y$ and between 28 and 106 in $z$.  The grid spacings were different.  The fine grid had a grid spacing of between $0.12$ and $0.25$ while the coarse grid always had a spacing which was three times larger.  The grids were placed so that they had a common center.   The fine grid lay entirely within the coarse grid and all calculations in the overlap were done using the fine grid, and then imposed upon the coarse grid at each step.

The fine grid spacing is similar to the $0.2$ used in Ref.~\cite{vachrel}.  However in that paper all fields reached their asymptotic values exponentially and so boundary conditions were not so essential.  In the present note, as $\lambda_1=0$, the $\Phi_1$ field falls as $1/r$ and so is still appreciable at the boundary of the fine grid.   This is the reason that we introduced the coarse grid.  Its boundaries are sufficiently distant that the $\Phi_1$ field approaches its asymptotic value.  However, in our simulations we keep the monopole centers well within the fine grid.  In fact we observed that, as a result of finite lattice spacing errors in the coarse grid, the monopoles are actually repelled from the coarse grid and so, unless they have a sufficiently high initial velocity, they do not leave the fine grid.

The time spacing in each simulation was always half of the spatial spacing in the fine grid.  We found that a larger time step leads to the numerical instabilities expected in the iterative Crank-Nicholson method when the time and spatial discretization scales are comparable.

\subsection{Results}

\begin{figure} 
\begin{center}
\includegraphics[width=4.0in,height=2.6in]{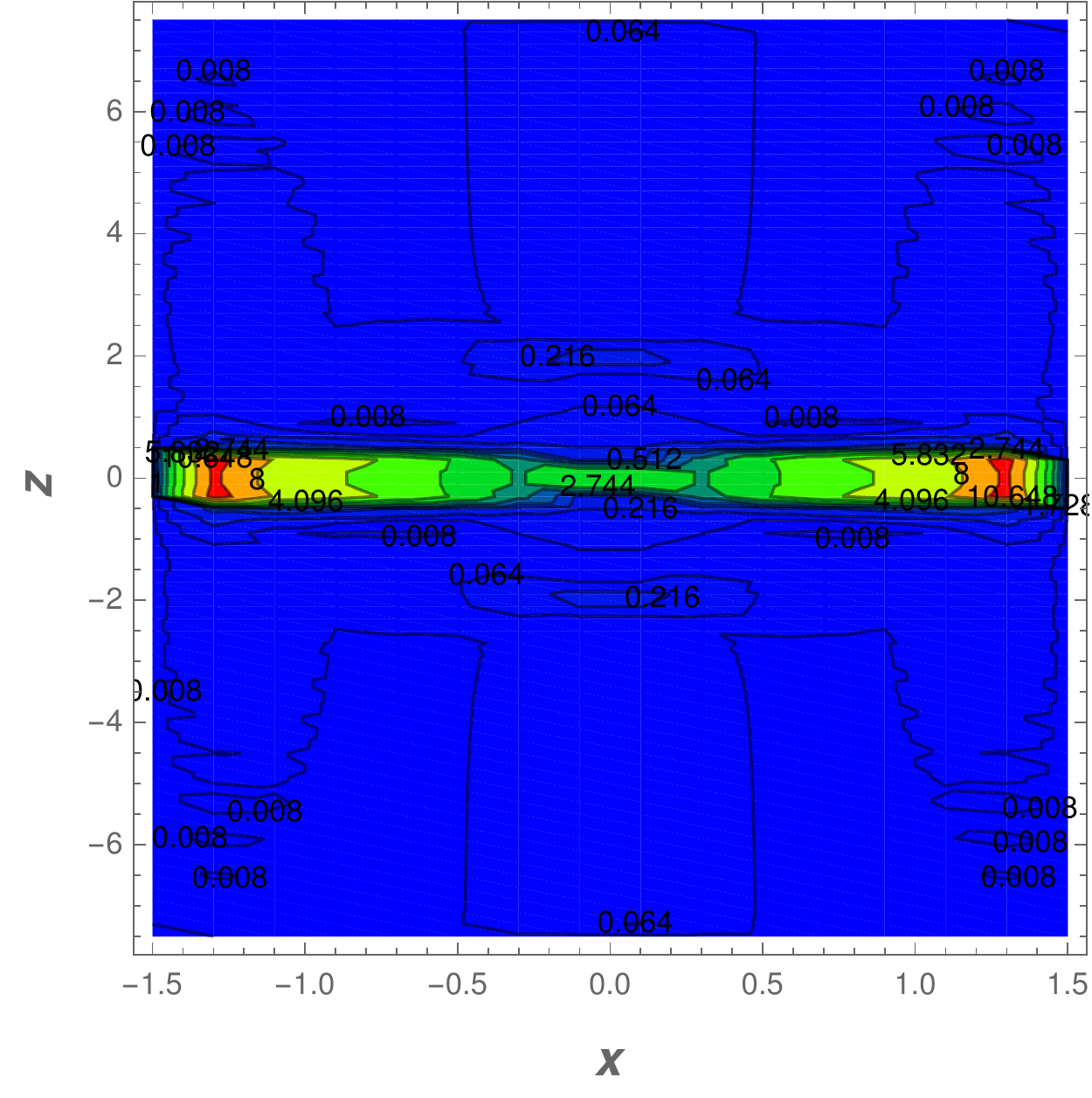}
\includegraphics[width=4.0in,height=2.6in]{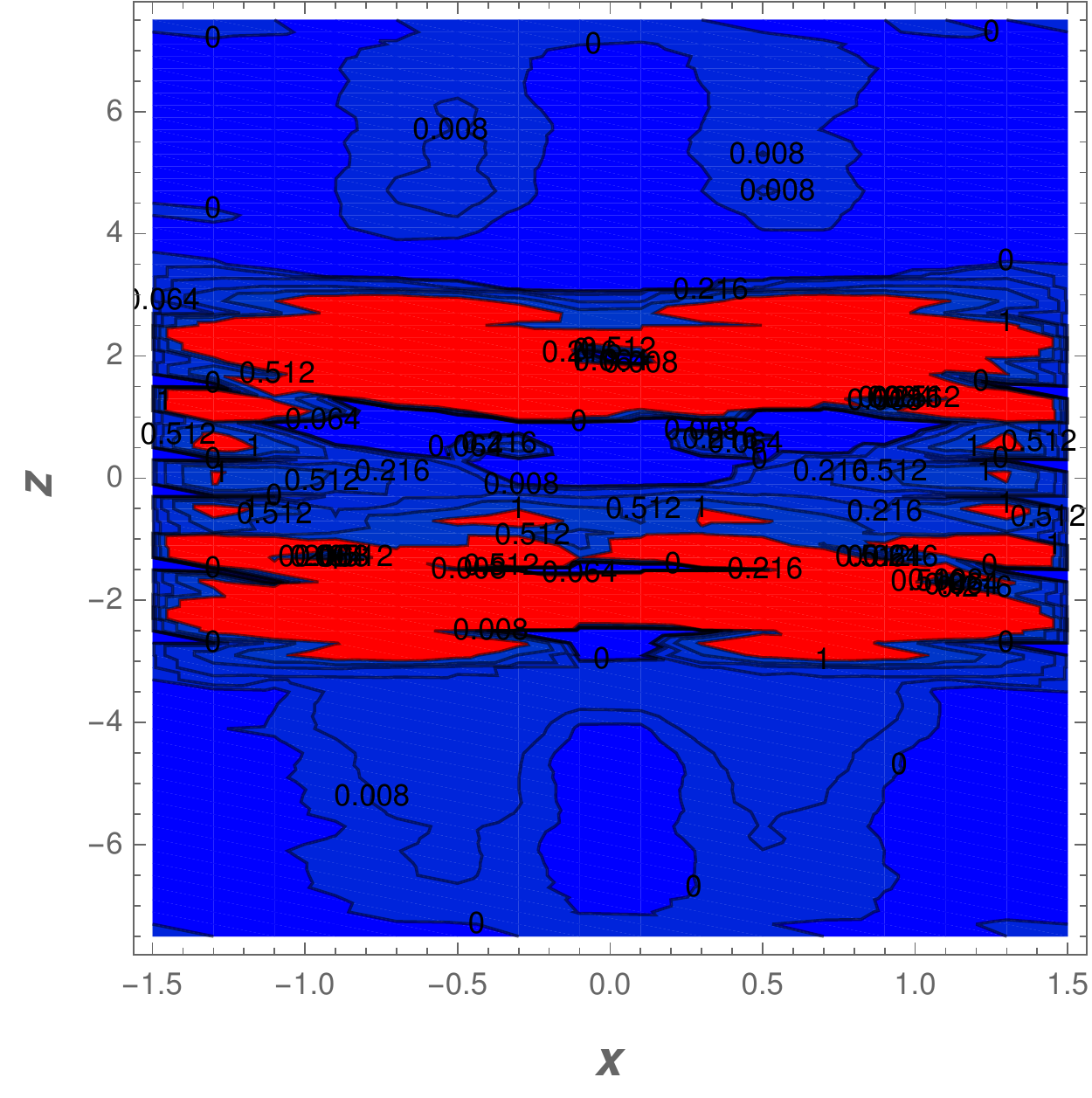}
\includegraphics[width=4.0in,height=2.6in]{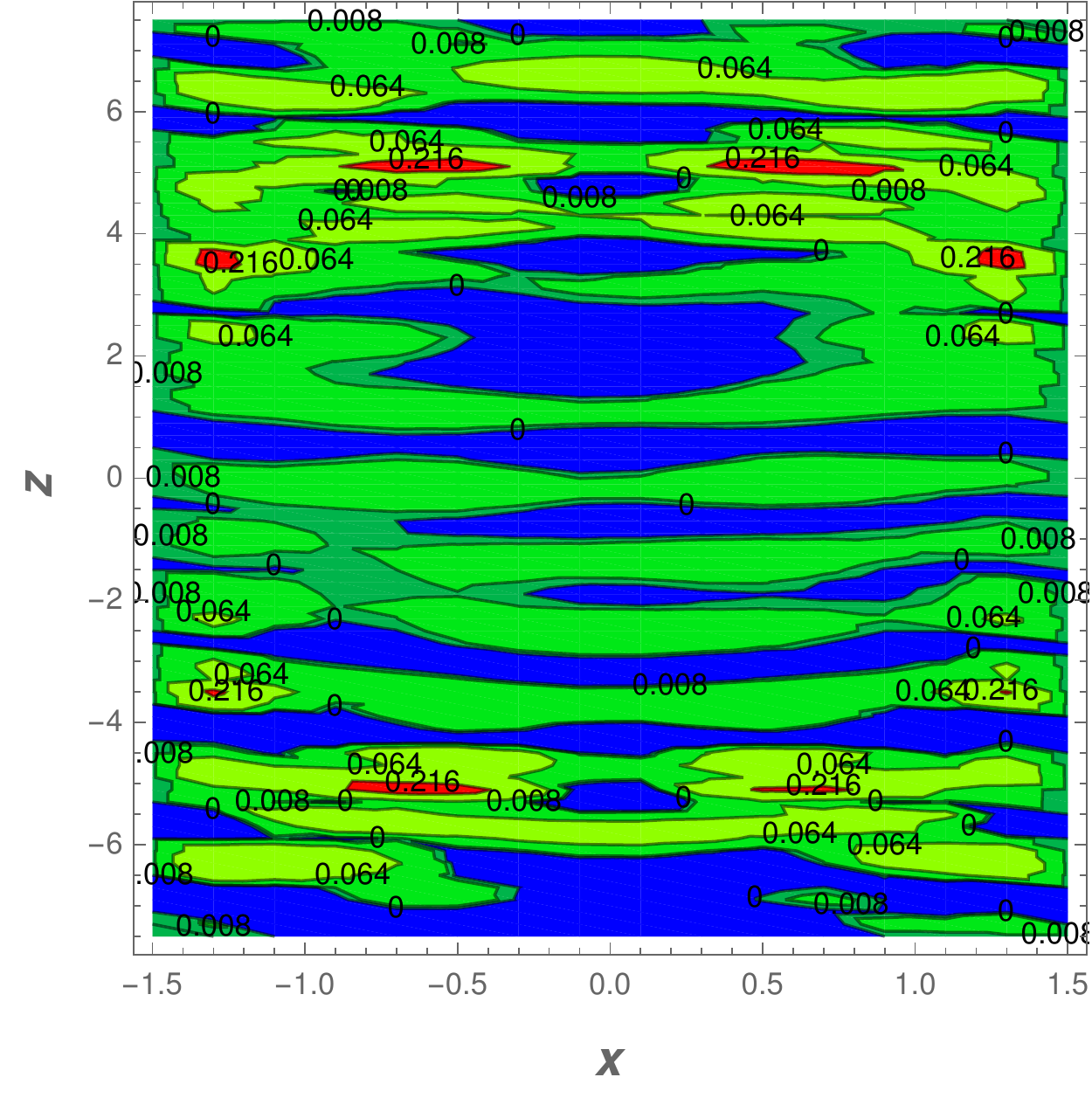}
\caption{The square root of the three components of the constraint added in quadrature at times $t=0.2$ (top), $t=3$ (middle) and $t=15$ (bottom) for two monopoles starting at $z=\pm 2$.  An initial violation of the constraint at $z=0$ spreads into the volume and diffuses as the system relaxes to a solution.  The finely spaced grid is plotted.}
\label{vincfig}
\end{center}
\end{figure}

In our two spiked monopole simulations, at $t=0$ the configuration is static and there is no electric field.  Therefore the gauge constraints are satisfied exactly.  However, our initial conditions are made by patching together two solutions, one at $z>0$ and the other at $z<0$.  This patching is imperfect, especially for small $z_0$.  As a result, there are violent derivatives near $z=0$ which, due to discretization errors, lead to an evolution which soon violates the constraints near $z=0$.  This violation spreads throughout the grid but eventually dissipates as Vachaspati's auxilliary field forces the configuration to relax back to a solution.  This relaxation is shown in Fig.~\ref{vincfig}.

\begin{figure} 
\begin{center}
\includegraphics[width=3.2in,height=2.2in]{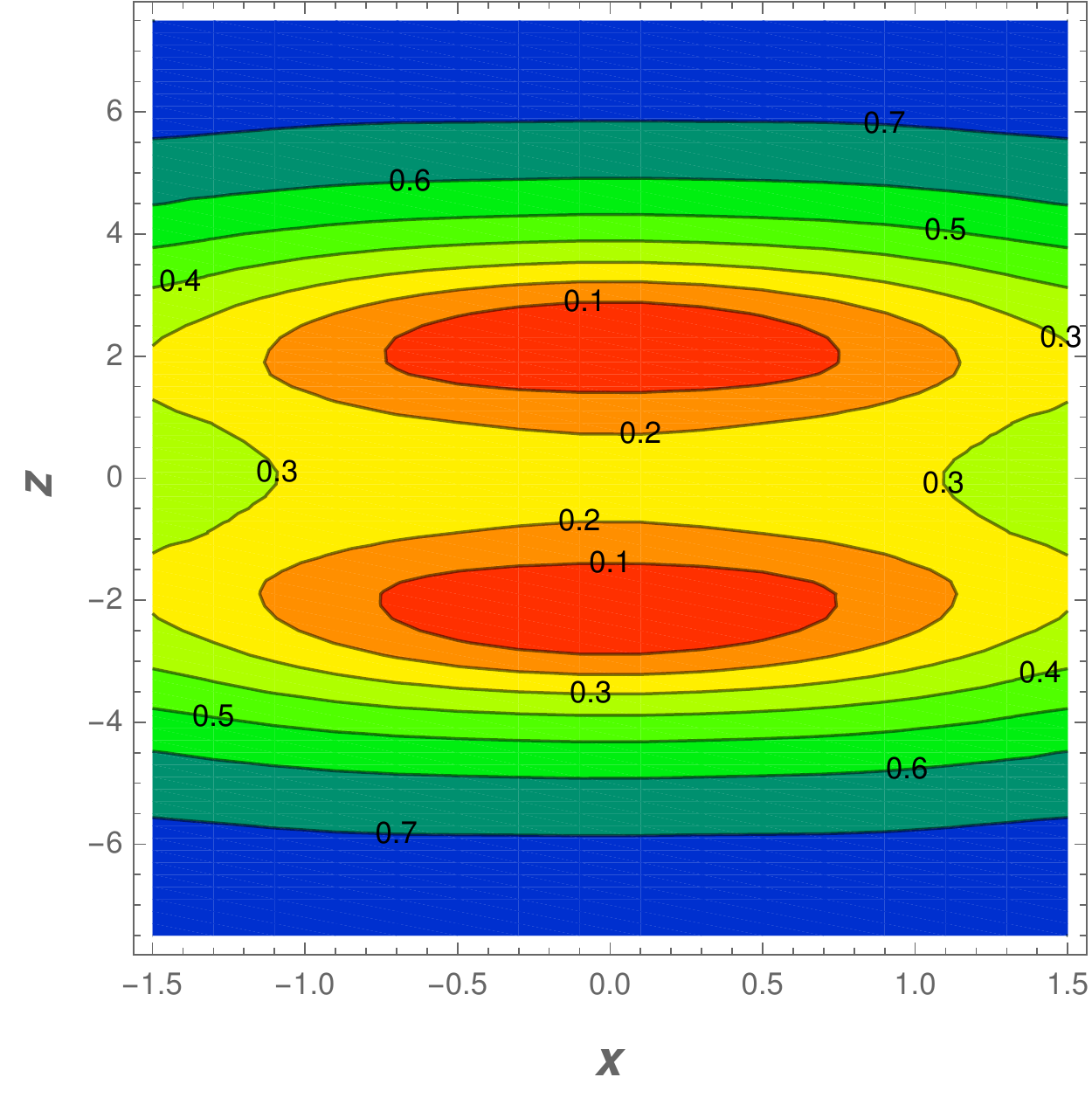}
\includegraphics[width=3.2in,height=2.2in]{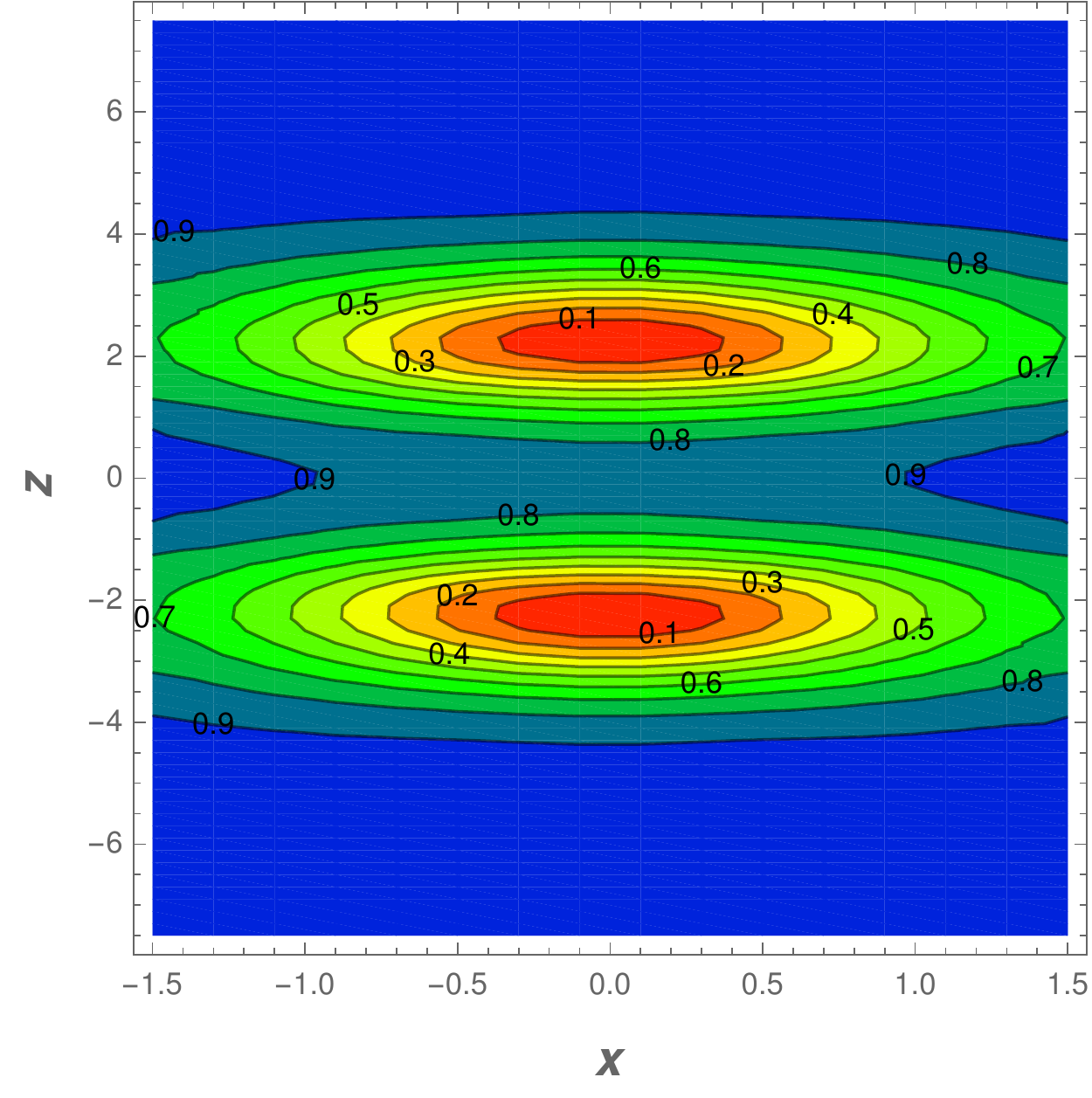}
\includegraphics[width=3.2in,height=2.2in]{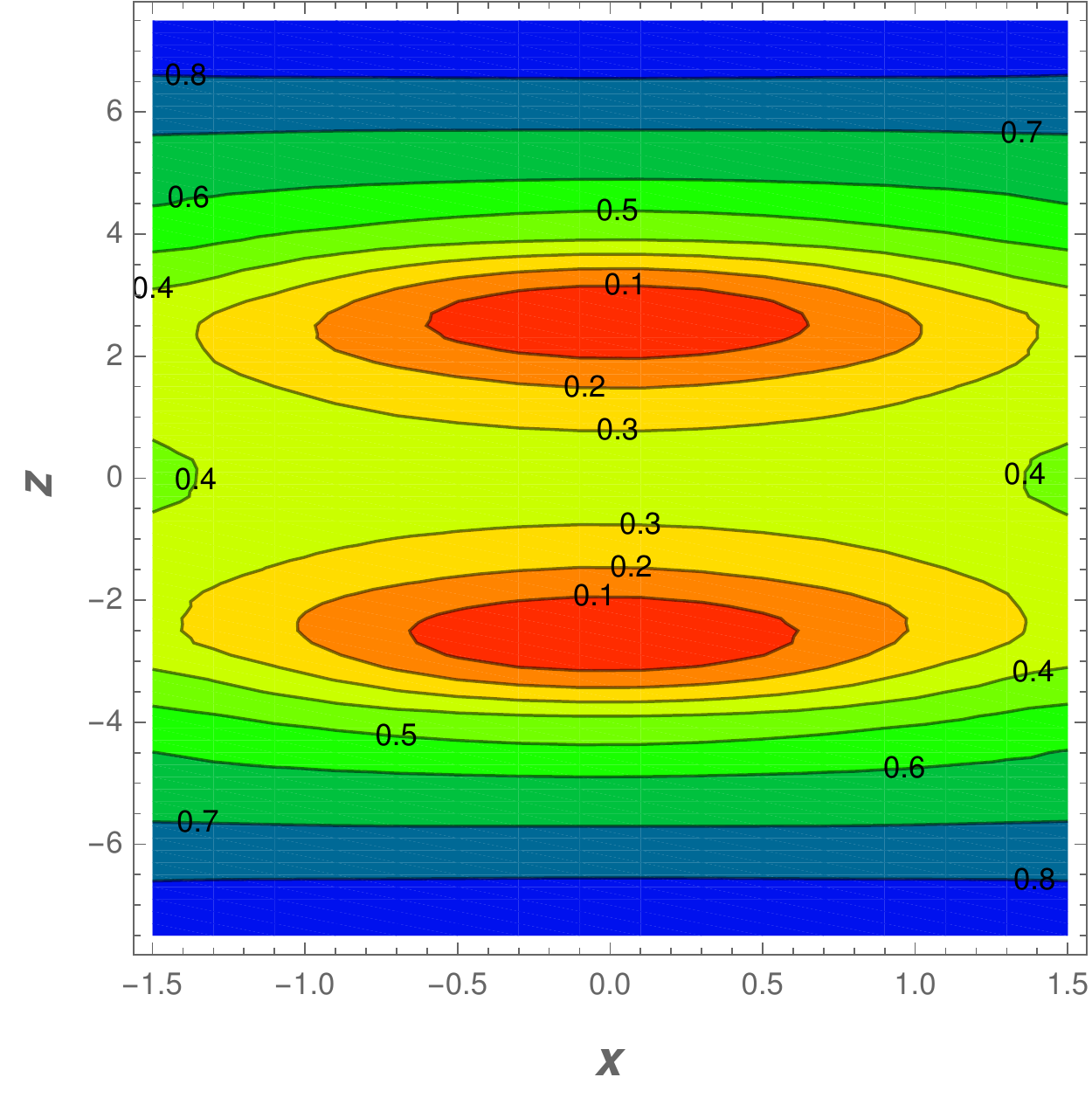}
\includegraphics[width=3.2in,height=2.2in]{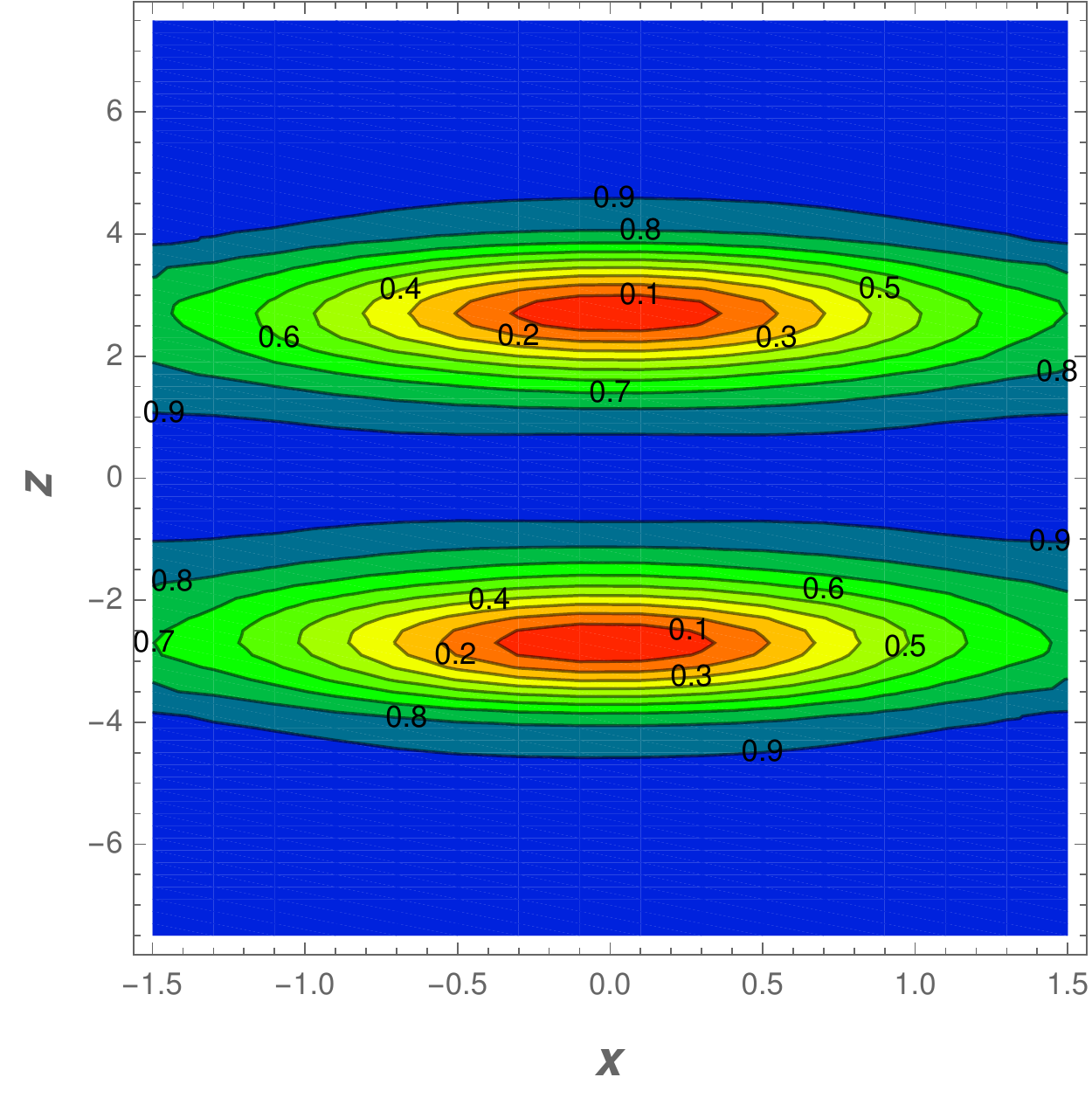}
\includegraphics[width=3.2in,height=2.2in]{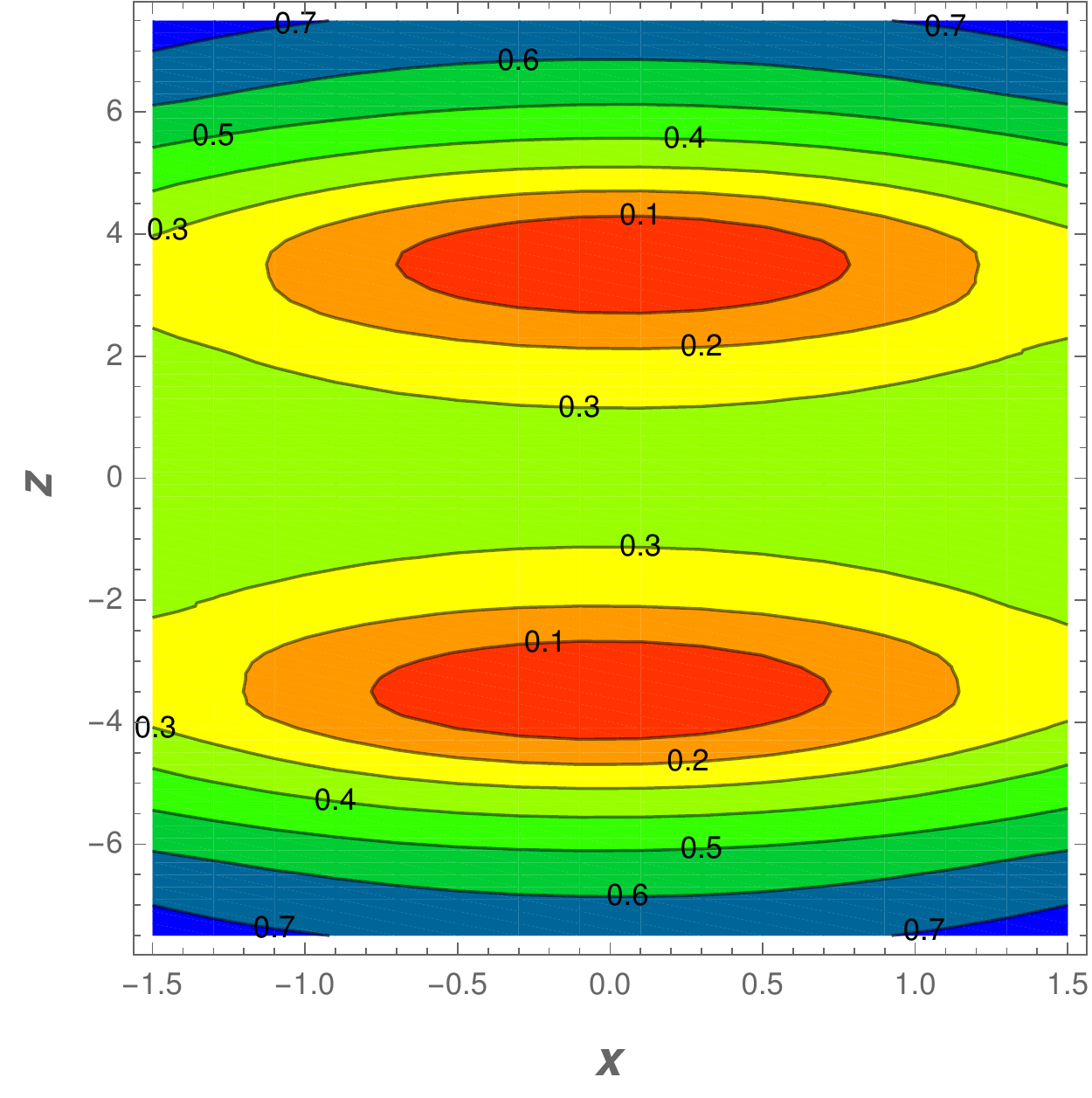}
\includegraphics[width=3.2in,height=2.2in]{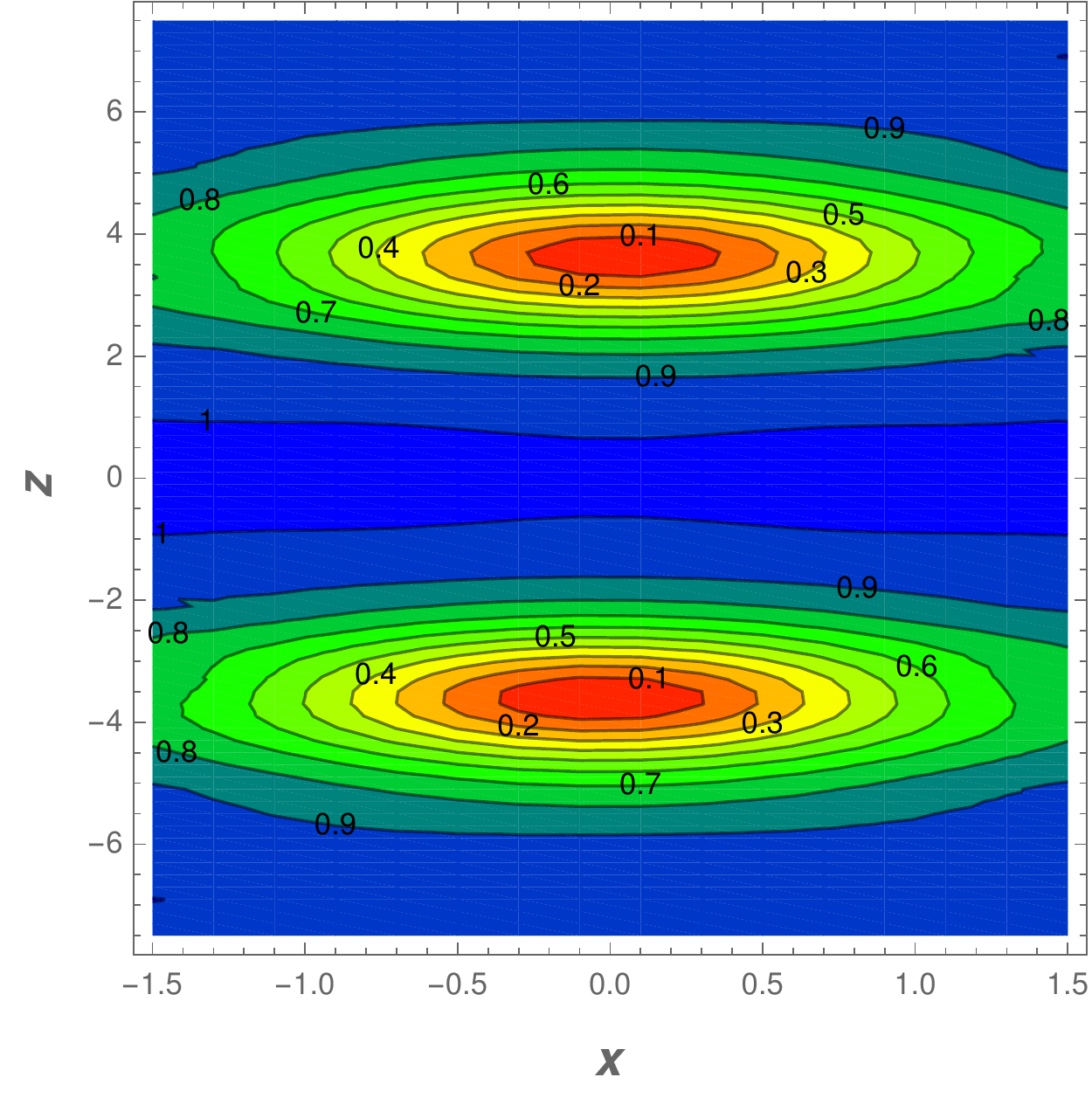}
\caption{The sum in quadrature of the three gauge components of scalar field $I=1$ (left) and $I=2$ (right) at $t=3$ (top), $t=6$ (middle) and $t=15$ (bottom).  The finely spaced grid is plotted.}
\label{phifig}
\end{center}
\end{figure}

We have observed that the scalar fields in the monopoles are remarkably stable, given initial velocities or when exposed to interactions with other monopoles.  The solutions change, but gauge-invariant observables move with the monopoles.  In Fig.~\ref{phifig} one can observe the motion of two monopoles beginning at $z=\pm 2$ on the finely spaced grid.  The self-interacting scalar field is tightly confined, yet it moves roughly in step with the other scalar field.   As expected, the monopoles repel.  In the last panel, one may see that on the two edges at $z=0$ the scalar field has actually decreased.    In fact, it fluctuates due to an effective friction force caused by numerical errors on the coarse grid.  Most of our simulations were done on wider grids where this effect is still present, but smaller.

\section{Comments}

Dark matter halos grow by merging.  This merging requires them to be attractive, but the simplest manifestation of monopole dark matter is repulsive.     If the magnetic repulsion is sufficiently weak, then it can be overcome by gravity.  However fitting parameters in the simplest model \cite{bjarkedark} one finds that $v\sim 10^{14}$\ GeV and so the magnetic repulsion is stronger than gravitational attraction by nearly 10 orders of magnitude.   In the spiked monopole model, the gravitational repulsion is reduced.  The crudeness of our numerical simulations and initial conditions makes it difficult to quantify the repulsion, however it clearly is not reduced by the required 10 orders of magnitude.  

When the mass in the spike is reduced, the asymptotic value of $\phi_1$ approaches its BPS value roughly linearly, not exponentially.  Therefore the attraction of the scalar field can cancel the repulsion with sufficient precision only if the mass of the spike is several orders of magnitude smaller than that of the monopole.  If it is the spike which yields flat galactic rotation curves, then it is not possible for the spike to be more than a few orders of magnitude lighter than its host halo.   Therefore we conclude that, without additional screening, it is not possible for this spike to explain the $1/r^2$ density profile of dark matter halos.  However, a much lighter spike which seeds black holes and perhaps resolves the final parsec problem \cite{yufinal} in their merging is allowed if the monopoles are either BPS or else screened by some other mechanism.

In a sequel we will attempt to examine the possibility that each halo consists of a gas of spiked BPS monopoles, whose spikes are small enough to induce little repulsion but large enough to seed black hole growth.  Each monopole in this gas should be small enough to evade the bounds in \cite{afshordishot}.  In principle, the entire monopoles themselves could serve as seeds, in which case the spikes are not necessary.  As the interactions between BPS monopoles depend on relative velocities and displacement \cite{gm} such a gas is rather complicated and its stability is still an open question, let alone its phenomenological viability as a halo model.

\section* {Acknowledgements}

\noindent
We thank Kimyeong Lee, Stefano Bolognesi and Paul Sutcliffe for invaluable comments which contributed to this project.  JE is supported by NSFC grant 11375201 and the CAS Key Research Program of Frontier Sciences grant QYZDY-SSW-SLH006.  JE also thanks the Recruitment Program of High-end Foreign Experts for support.


\end{document}